# Grain Boundary Motion on Curved Substrate

Kongtao Chen

*Department of Materials Science and NanoEngineering, Rice University, Houston, TX 77005 USA*

## Abstract

Grain boundary (GB) kinetics is important for many applications in 2d materials and metal thin films. To study how the substrate shape affects GB mobility and kinetics, we develop a kinetic Monte-Carlo (kMC) simulation method and an analytical model for GBs on the curved substrate by combining disconnection theory and by Föppl–von Kármán equations. Using sinusoidal $MoS_2$ as an example, we can increase its GB mobility more than 50 times by changing substrate shape amplitudes and periods. We find that amplitude change GB mobility exponentially while wave vector change GB mobility linearly. The sinusoidal GB kinetic shape has wave vector twice as substrate and amplitude proportional to substrate squared amplitude.

## Introduction

Grain boundary (GB) kinetics is important for many applications in 2d materials[1] and metal thin films[2]. GB motions are used to absorb stacking faults and rotational disorders in 2d transition metal dichalcogenides (TMDs)[3]. Polycrystalline molybdenum disulfide ($MoS_2$) cracks along GB[4] if GBs don't move and stress accumulates. GB orientations, which affect mechanical properties[5], failure modes[1], electric transport[6] of 2d TMDs, may change when GB migrates. GB motions are also observed in polycrystalline $MoS_2$ memtransistors[7]. The single-crystal metal substrate useful for the growth of single-crystal TMDs is often processed by grain growth[2]. The dynamics of flat GBs in 2d materials have been studied by dislocation dynamics[8–15] and molecular dynamics simulations[16]. How substrate shape affects GB dynamics, however, hasn't been studied yet.

A disconnection-based model[17–24] for GB motion in bulk materials was recently developed to systematically and quantitatively explain many GB properties, e.g., shear coupling[18–20], mobility[20,21], sliding[20], roughening[22] and phase transitions[22]. This disconnection-based model was recently applied on GBs in 2d materials[25,26], which was traditionally regarded as arrays of dislocations[1].

We develop a disconnection-based kinetic Monte-Carlo (kMC) simulation method for GBs on curved substrate to study how the substrate shape affects GB mobility and kinetic shapes. The stress and strain fields are calculated by Föppl–von Kármán equations[27–29].

## Methods

We first determine the stress and strain fields from Föppl–von Kármán equations[27–29], and then perform the kMC simulations proposed by Chen et al.[21] Assume the 2d material (and substrate) has a periodic shape

$$z = \frac{c}{\alpha} \sin \alpha x \sin \alpha y \qquad (1)$$

Where $x$, $y$, and $z$ are defined in Figure 1. When strain is small ($c \ll 1$), Föppl–von Kármán equations[27–29] leads to strain field

$$\epsilon_{xx} = \frac{c^2}{8}(\nu \cos 2\alpha x - \cos 2\alpha y) \qquad (2)$$

$$\epsilon_{yy} = \frac{c^2}{8}(\nu \cos 2\alpha y - \cos 2\alpha x) \qquad (3)$$

Where $\nu$ is Poisson's ratio. The shear stress $\sigma_{xy} = 0$. The disconnection Burgers vector $b$ and step height $h$ at different locations will rescale with $\epsilon_{xx}$ and $\epsilon_{yy}$, respectively.

In the kMC simulation (See Figure 1), each event corresponds to changes in the disconnections of mode $m$ at site $i$; the associated barrier $\Delta E_{im}$ is determined by the constant glide barrier and the change in energy of the state after the occurrence of the event. The rate of such an event is proportional to $exp(-\Delta E_{im}/k_B T)$. At each kMC step, we randomly choose one event from a list of all possible events (weighted by their rates), execute that event, and advance the clock in accordance with the sum of the rate of all possible events. GB motion is driven by a chemical potential jump $\psi$ across GB. A more detailed description of the kinetic Monte Carlo method is in Chen et al.[21] We set GB energy $\gamma = 1\,J/m^2$, temperature $T = 1500$ K, and use the material parameters of $MoS_2$ (lattice constant $a = 0.32$ nm, thickness $w = 1.23$ nm, shear modulus $G = 50$ GPa, Poisson's ratio $\nu = 0.25$)[30–32].

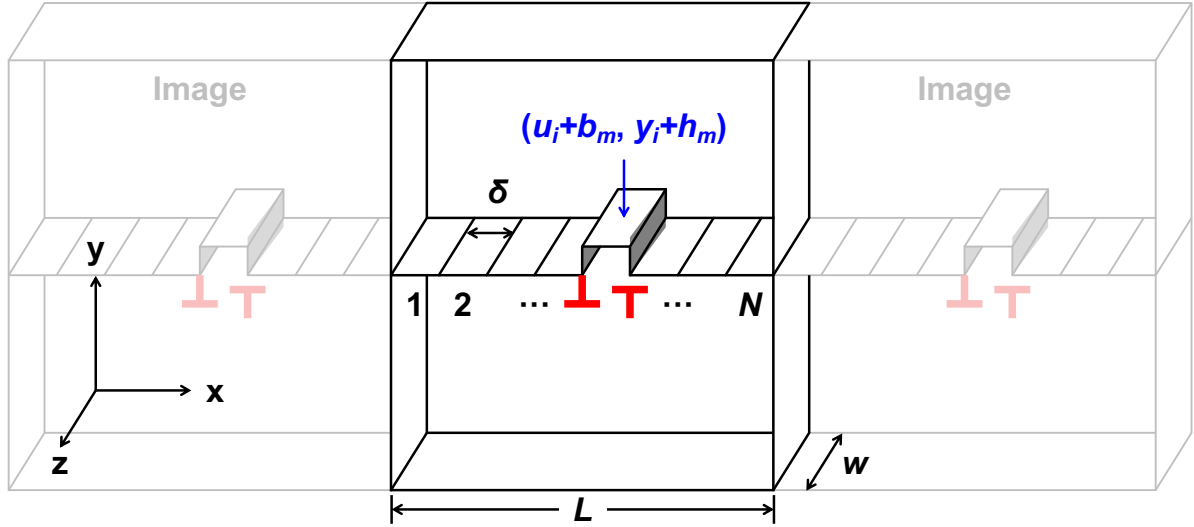

*Figure 1 Disconnection dipoles on a 1d GB in 2d materials. $\delta$, w, and L denote the disconnection core size, thickness of the material, and the system size in the y-direction, respectively. The states of the system at i before and after disconnection dipole nucleation are $(u_i, z_i)$ and $(u_i + b_m, z_i + h_m)$, respectively, for a pair of disconnections of mode m at location i.*

## Results

Figure 2 shows that the time-averaged kinetic GB shape when migration is a sinusoidal function

$$y \propto c^2 h \sin 2\alpha x \qquad (4)$$

The wave vector is the same as $\epsilon_{xx}$ and $\epsilon_{yy}$ (or twice as substrate). This is because $2\alpha$ is the wave vector of the whole periodic simulation system. The amplitude $D \propto c^2 h$, also proportional to $\epsilon_{xx}$ and $\epsilon_{yy}$.

Figure 3 shows that reduced mobility $\widetilde{M}$ (the GB mobility on a substrate described by Equation (1) over the mobility of a flat GB) is proportional to substrate wave vector $\alpha$, and $\ln \widetilde{M}$ is proportional to $c$. This is because the disconnection nucleation barrier per thickness at any site

$$\frac{Q}{w} = Ab^2(1 + \epsilon_{xx})^2 + 2\gamma|h(1 + \epsilon_{yy})| + C \tag{5}$$

has $N\alpha/\pi$ minima, $Q_{min}/w = Ab^2[1 - c^2(1+v)/8]^2 + 2\gamma|h(1 - c^2(1+v)/8)|$, where $A = -2G[(1 - v\cos^2\beta)/4\pi(1-v)] \ln \sin(\pi r_0/L)$, $G$ is the shear modulus, $v$ is the Poisson's ratio, $\beta$ is the angle between the Burgers vector and the disconnection line direction, and $r_0$ is the disconnection core size. $A$ describes the energy required to form a dislocation pair and separate it to a distance of half the periodic unit cell $L/2$ and GB energy $\gamma$ describes the energy required to form a pair of steps[17–23]. $C$ represents the disconnection migration barrier which depends on the GB structure and bonding character; this is dominated by core-level phenomena and may be determined via calculations on the atomic scale[20,33]. Assume disconnections only nucleate at these minimum sites and bulk material parameters ($A$, $\gamma$, $C$) don't change under small strain, the reduced mobility

$$\widetilde{M} = \frac{M(\alpha, c)}{M(\alpha = 0, c = 0)} = \frac{N\alpha}{\pi} \exp\left[\frac{(Ab^2 + 2\gamma h)c^2(1+v)w}{4k_BT} + O(c^4)\right] \tag{6}$$

This result is consistent with simulation results in Figure 3.

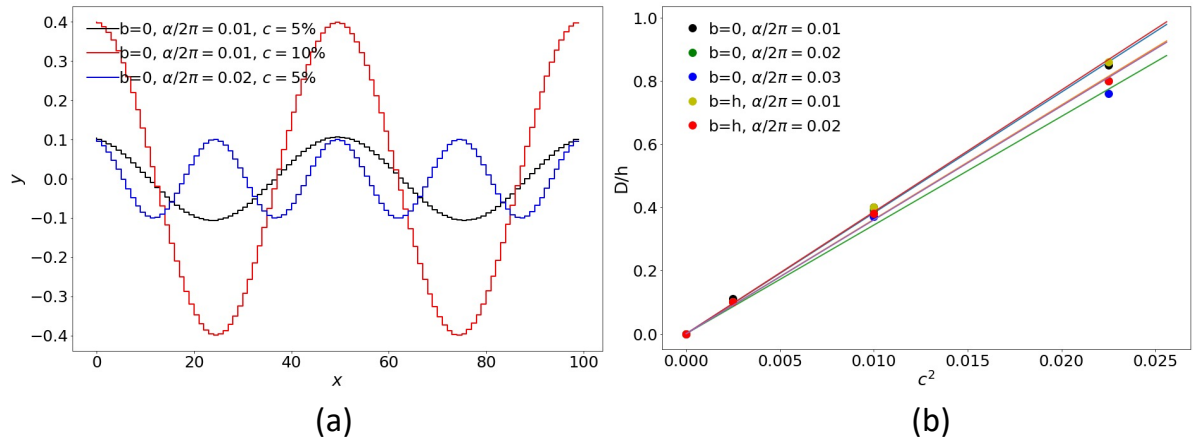

*Figure 2 Kinetic GB shapes on curved 2d materials. The time-averaged kinetic GB shape when migration is a sinusoidal function with twice the same wave vector as substrate and amplitude $D \propto c^2$.*

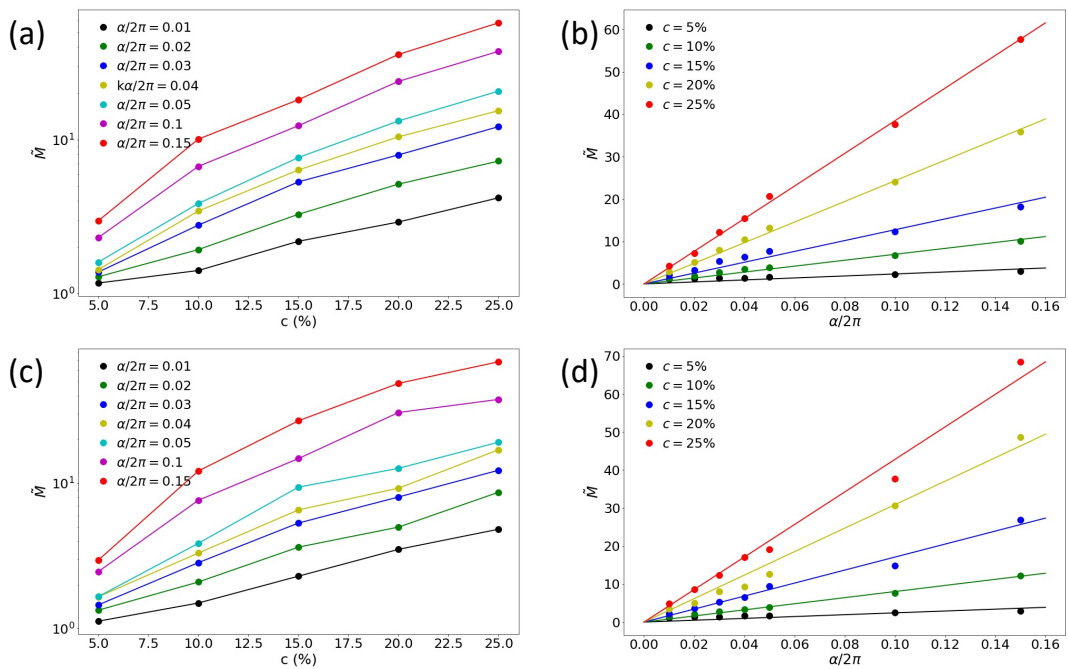

Figure 3 GB mobility as a function of substrate shape. Reduced mobility $\tilde{M}$ is the GB mobility on a substrate described by Equation (1) over the mobility of a flat GB. The disconnection mode in (a) and (b) is $b = 0$, $h = a$; the mode in (c) and (d) is $b = h = a$.

# Conclusion

We develop a kMC simulation method and analytical model for GB kinetics of 2d materials on non-uniform substrates, by combining disconnection theory and by Föppl–von Kármán equations. Using $MoS_2$ as an example, we can increase its GB mobility for $> 50$ times by changing substrate shape amplitudes and periods. We find that amplitude change GB mobility exponentially while wave vector change GB mobility linearly. The sinusoidal GB kinetic shape has wave vector twice as substrate and amplitude proportional to substrate squared amplitude.